\documentclass [twocolumn,pra,showpacs,preprintnumbers,amsmath,amssymb,showkeys]{revtex4}

\usepackage{graphicx}
\usepackage{dcolumn}
\usepackage{bm}

\begin{document}

\preprint{APS/123-QED}

\bibliographystyle{apsrev}
\title{Nonlinear interaction of light with Bose-Einstein condensate:
new methods to generate subpoissonian light}
\author{S.~M.~Arakelyan}
\email{avprokhorov@front.ru} \affiliation{Vladimir State
University, ul. Gor'kogo 87, Vladimir, 600000 Russia}
\author{A.~V.~Prokhorov}
\email{avprokhorov@front.ru} \affiliation{Vladimir State University,
ul. Gor'kogo 87, Vladimir, 600000 Russia}
\author{I.~Vadeiko}
\email{iv3@st-andrews.ac.uk} \affiliation{School of Physics and
Astronomy, University of St Andrews, North Haugh, St Andrews, KY16
9SS, Scotland}
\date{\today}
\begin{abstract}
We consider $\Lambda$-type model of the Bose-Einstein condensate of
sodium atoms interacting with the light. Coefficients of the
Kerr-nonlinearity in the condensate can achieve large and negative
values providing the possibility for effective control of group
velocity and dispersion of the probe pulse. We find a regime when
the observation of the "slow" and "fast" light propagating without
absorption becomes achievable due to strong nonlinearity. An
effective two-level quantum model of the system is derived and
studied based on the su(2) polynomial deformation approach. We
propose an efficient way for generation of subpoissonian fields in
the Bose-Einstein condensate at time-scales much shorter than the
characteristic decay time in the system. We show that the quantum
properties of the probe pulse can be controlled in BEC by the
classical coupling field.
\end{abstract}
\maketitle

\section{Introduction}
The problem of atom-field interaction represent one of the major
areas in the modern physics and quantum optics, particularly
\cite{Scully:1997, Wolf:2002}. A special interest is regarded to
effects of propagation of light inside highly nonlinear atomic media
where a large nonlinearity is achieved with appropriate choice of
the form of the atom-field interaction \cite{Imamoglu:1997,
Wang:2002}. Some early works \cite{Basov:1966, Let:1966} on
interaction of a resonance atomic system and laser field
demonstrated a possibility to achieve significant difference between
phase and group velocity of the light pulses. In the experiments,
Letokhov and Basov showed that while the front-edge of the pulse
generated the inversion in the system of two-level particles
resulted in the sloping of the front-edge, the back process of
re-emission of the absorbed energy gave rise to some steeping of the
back-edge of the pulse. Hence, the pulse shape deformation resulted
in significant delay of its registration on exit from the resonance
media that was observed experimentally. The works motivated
intensive study of so-called the "slow" and "fast" light.

An important step in the development of the theory was the
theoretical prediction \cite{Garrett:1970} and experimental
observation \cite{Wong:1982} of positive and negative delays of
picosecond pulses propagating without pulse shape deformation inside
a crystal. But, very high level of optical losses was limiting the
effect to a relatively small magnitude. Basically, there are several
different ways to overcome the difficulty. The first idea exploits
the shortening of the pulse duration to lengths, which would be much
smaller than the relaxation times in the medium that provides
necessary conditions for a generation of optical solitons
\cite{Allen:1978}. Another approach is based on a three-level
$\Lambda$-scheme. In that case one of two laser beams is a strong
coupling field developing a transparency window in the medium and
the second probe field propagates through the resonant system
without absorption and with unchanged pulse shape
\cite{Harris:1990}.

\begin{figure}
\includegraphics[width=60mm]{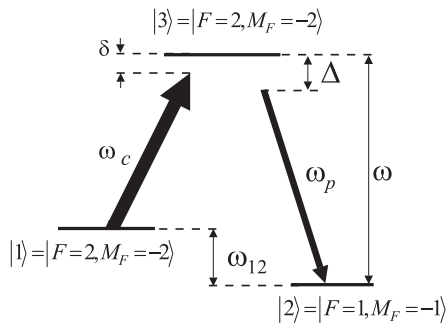}
\caption{\label{fig:spec1} The energy level $\Lambda$-scheme of
$^{23}Na$ atoms.}
\end{figure}

A sketch of the energy levels in the $\Lambda$-scheme for a single
sodium atom is shown in Fig.~\ref{fig:spec1}. A classical beam with
the central frequency $\omega_c$ couples the levels $|1\rangle$ and
$|3\rangle$, and the probe pulse with central frequency $\omega_p$
couples $|3\rangle$ and $|2\rangle$ levels such that the three
levels form a $\Lambda$-type configuration. If the intensities of
the classical and probe fields are comparable and the relaxation
rates of the levels $|1\rangle$ and $|2\rangle$ are negligible then
a state of an atom may be considered as quantum superposition of two
lower levels $|1\rangle$ and $|2\rangle$. It corresponds to the
effect of so-called coherent trapping of lower states
\cite{Alzetta:1976,Gray:1978}. Another type of dynamics may be
realized if the intensity of the probe pulse is much smaller than
the intensity of the classical field and the atoms initially
populate the lower level $|2\rangle$. The regime is called
electromagnetically induced transparency (EIT). In the paper we
concentrate on the later.

Systems in the EIT regime are extensively studied in the literature.
Some interesting results are referred to: the studies of the
dispersion of the atomic medium in a linear response regime
\cite{Agarwal:1986}, semiclassical estimates of the time delay of
the probe field inside atomic cells in a thermal equilibrium
\cite{Harris:1992}, EIT in doped crystals \cite{Turukhin:2002},
nonlinear optical parametric processes in resonant double-$\Lambda$
system \cite{Lukin:1998}. Wang and colleagues carried out an
important experimental measurements of the Kerr nonlinear index of
refraction in a three-level Rb atomic $\Lambda$-system using an
optical ring-cavity \cite{Wang:2001}. They found that the Kerr
nonlinearity can reach very large and negative values.

One of the dominating problems in EIT was the spectrum broadening
regarded to thermal effects. To overcome the difficulty, a highly
coherent atomic Bose-Einstein condensate (BEC) is used
\cite{Ketterle:2002,Cornell:2002}. Hau and co-workers calculated
linear dispersive properties of the system and demonstrated a
possibility to observe the "slow" light effect in the ultra-cold
vapor of sodium atoms loaded into a magnetic trap  at the
temperature of transition to the condensed state \cite{Hau:1999}.
Based on their results we consider EIT  effects in the Bose-Einstein
condensate of sodium atoms. Two hyperfine sub-levels of sodium state
$3^2S_{1/2}$ with $F=1, F=2$ are associated with levels $|2\rangle$
and $|1\rangle$ of the $\Lambda$-scheme, correspondingly. An excited
state $|3\rangle$ corresponds to the hyperfine sub-level of the term
$3^2P_{3/2}$ with $F=2$ (see Fig.~\ref{fig:spec1})
\cite{Kelley:1985}. The energy splitting between the levels
$|1\rangle$ and $|2\rangle$ is denoted by $\omega_{12}/(2\pi)=1772
\mathrm{MHz}$, the transition $|3\rangle \rightarrow |2\rangle$
corresponds to the optical frequency $\omega/(2\pi)=5.1\cdot 10^{14}
\mathrm{Hz}$ ($\lambda=589 \mathrm{nm}$) \cite{Mitsunaga:2000}.

The condensate is considered being placed inside a confocal
resonator. A strong coupling beam propagates along the optical axis
of the resonator maintaining the transparency window in the
condensate due to the transitions $|1\rangle \rightarrow |3\rangle$.
According to the conventional EIT regime the atoms mainly populate
the lowest level $|2\rangle$. We assume that the resonator mode has
an eigenfrequency $\omega_c$. The coupling laser beam develops large
polarization in the condensate resulting in significant nonlinear
susceptibility of the medium on the frequency $\omega_p$. A probe
pulse with the appropriately chosen polarization direction
propagates transversely to the resonator optical axis. It obeys high
nonlinear interaction with the condensate. In order to study a
competition between linear and nonlinear processes regarded to the
transition $|3\rangle \rightarrow |2\rangle$ induced by the probe
pulse, the linear and third order Kerr-like terms in the expansion
of atomic susceptibility are considered. An effective control on the
absorption of the probe pulse and its group velocity is realized by
detuning $\delta$ of the coupling field frequency from the
resonance.

In the following section we find conditions when the group velocity
of the probe pulse becomes very small or very large. We also note a
region where the absorption in the condensate is almost zero what we
have called as nonlinear compensation of optical losses. In the
third section, the interaction between sodium BEC and the probe
pulse is described by an effective quantum Hamiltonian. We explain
the way to reformulate linear three level $\Lambda$-scheme in terms
of nonlinear quantum model of a two-level particle interacting with
quantized field and to solve the model. We apply the su(2)
polynomial deformation formalism in order to develop the
perturbation theory. In the next section the perturbation theory is
considered up to second order. The last fifth section concerns with
some nonclassical effects in quantum statistics of the photons in
the probe pulse described by the effective Hamiltonian. In the
conclusion we discuss some experimental perspectives of our results.

\section{Nonlinear compensation in sodium BEC.}
\label{sec:part1}

Applying the formalism of slowly varying field amplitudes
\cite{Leontov:1944} in the rotating wave approximation we write the
Hamiltonian describing the interaction of three-level atoms with two
laser fields in the following form \cite{Patnaik:2002}:
\begin{eqnarray}
\label{H_int_lam} &H^\Lambda=\omega_{12}|1\rangle\langle1|+\omega
|3\rangle\langle3|\nonumber\\
&-g_1|3\rangle\langle1| e^{-i\omega_c
(t-\frac{z}{c})}-g_1^*|1\rangle\langle3| e^{i\omega_c
(t-\frac{z}{c})}
\nonumber\\
&-g_2|3\rangle\langle2| e^{-i\omega_p (t-\frac{z}{c})}-
g_2^*|2\rangle\langle3| e^{i\omega_p (t-\frac{z}{c})}.
\end{eqnarray}
For simplicity, we let $\hbar=1$ in the paper. In
Eq.~(\ref{H_int_lam}) the coefficients $g_{1,2}$ determine
single-photon Rabi frequencies and are defined as follows
\begin{eqnarray}
\label{coup_const} g_1=|\mu_{31}| A_c,\; g_2=|\mu_{32}| A_p.
\end{eqnarray}
Here, $\mu_{ij}$ is the atomic dipole momentum, $A_{c(p)}$ are the
slowly varying coupling and probe field amplitudes,
correspondingly.

We denote single-atom density matrix by
$\rho=\sum_{\{i,j\}=1}^3{\rho_{ij}|i\rangle\langle j|}$. The time
evolution of the density matrix is described by the Liouville
equation of motion  \cite{Shen:1984}
\begin{eqnarray}
\label{liouv}& \frac{\partial \rho}{\partial t}=-i
\left[{H^\Lambda,\rho}\right]\nonumber\\
&-\gamma_{12}(|1\rangle\langle 1|\rho-2\,|2\rangle\langle
1|\rho|1\rangle\langle 2|+\rho|1\rangle\langle 1|)
\nonumber\\
&-\gamma_{32}(|3\rangle\langle 3|\rho-2\,|2\rangle\langle
3|\rho|3\rangle\langle 2|+\rho|3\rangle\langle 3|)
\nonumber\\
&-\gamma_{31}(|3\rangle\langle 3|\rho-2\,|1\rangle\langle
3|\rho|3\rangle\langle 1|+\rho|3\rangle\langle 3|).
\end{eqnarray}
Here, the constants $\gamma_{ij}$ determine the rate of spontaneous
decay from the levels $|i\rangle$ to $|j\rangle$ in the
$\Lambda$-scheme. In general, to consider the space-time dynamics of
the system of fields and atoms interacting in the resonator we have
to add to Eq.~(\ref{liouv}) the Maxwell equations
\begin{equation}
\label{fields_eq}
\bigtriangledown\times\bigtriangledown\times\vec{\mathbf{E}}=
-\frac{1}{c^2}\frac{\partial^2\vec{\mathbf{E}}}{\partial
t^2}-\frac{1}{\varepsilon_{0}c^2}\frac{\partial^2\vec{\mathbf{P}}}{\partial
t^2} . \tag{\ref{liouv}a}
\end{equation}
Here, $c$ is the velocity of light in the vacuum, $\vec{\mathbf{E}}$
is the total amplitude of corresponding fields. The vector
$\vec{\mathbf{P}}$ is the polarization of the condensate induced by
the coupling or probe field, correspondingly.

In the adiabatic limit, when the variation of the Rabi frequency
$g_1$ is very small \cite{Lukin:2000} a self-consistent problem of
Eqs.~(\ref{liouv}) and (\ref{fields_eq}) may be reduced to a less
complicated one. In that case the system of equationscan be solved
separately  for the medium and the fields. Averaged the density
matrix elements over the rapidly oscillating phase of the fields we
represent $\rho$ in the following form
\begin{eqnarray}
\label{rho_aver} &\rho_{ii}=\bar\rho_{ii},\;
\rho_{31}=\bar\rho_{31} e^{-i\omega_c (t-\frac{z}{c})},\;
\rho_{32}=\bar\rho_{32} e^{-i\omega_p (t-\frac{z}{c})},\nonumber\\
&\rho_{12}=\bar\rho_{12} e^{-i(\omega_p-\omega_c)
(t-\frac{z}{c})},
\end{eqnarray}
together with the relation $\bar\rho_{ij}=\bar\rho_{ji}^*$. A bar in
the matrix elements $\bar\rho$ denotes the averaged quantities.
Substituting definitions Eq.~(\ref{rho_aver}) into Eq.~(\ref{liouv})
one obtains equations of motion for $\bar\rho$ matrix elements
\begin{eqnarray}
\label{rho_aver_eq} &\dot{\bar\rho}_{11}=-ig_1\bar\rho_{13} +i
g_1^* \bar\rho_{31} - 2\gamma_{12}
\bar\rho_{11}+2\gamma_{31}\bar\rho_{33},\nonumber\\
&\dot{\bar\rho}_{22}=-ig_2\bar\rho_{23} +i g_2^* \bar\rho_{32} +
2\gamma_{12} \bar\rho_{11}+2\gamma_{32} \bar\rho_{33},\nonumber\\
&\dot{\bar\rho}_{33}=ig_1\bar\rho_{13} -i g_1^* \bar\rho_{31}
+ig_2\bar\rho_{23} -i g_2^* \bar\rho_{32}
-2(\gamma_{32}+\gamma_{31}) \bar\rho_{33},\nonumber\\
&\dot{\bar\rho}_{21}=-i(\delta-\Delta) \bar\rho_{21}-i g_1
\bar\rho_{23} +ig_2^* \bar\rho_{31}
-\gamma_{12}\bar\rho_{21},\\
&\dot{\bar\rho}_{31}=-i\delta \bar\rho_{31}+i g_1 (\bar\rho_{11}-
\bar\rho_{33}) + ig_2 \bar\rho_{21} \nonumber\\
&-(\gamma_{12}+\gamma_{32}+\gamma_{31})\bar\rho_{31},\nonumber\\
&\dot{\bar\rho}_{32}=-i\Delta \bar\rho_{32} +i g_2
(\bar\rho_{22}-\bar\rho_{33}) +ig_1 \bar\rho_{12}
-(\gamma_{32}+\gamma_{31})\bar\rho_{32}.\nonumber
\end{eqnarray}
Here, $\Delta=\omega-\omega_p$ and $\delta=\omega-
\omega_{12}-\omega_c$.

Since we are interested in nonlinear interaction of the BEC with the
probe pulse in the dipole approximation we only consider the
explicit dependence of $\bar\rho_{32}$ on the Rabi frequency $g_2$
of the probe field,
\begin{eqnarray}
\label{rho_nonlin}
\bar\rho_{32}=\bar\rho_{32}^{(0)}+\bar\rho_{32}^{(1)}
g_2+\bar\rho_{32}^{(2)} |g_2|^2+\bar\rho_{32}^{(3)} |g_2|^2 g_2,
\end{eqnarray}
where $\bar\rho_{32}^{(0)}=0$ denotes initial polarization in the
condensate, which is zero for the sodium. The coefficient
$\bar\rho_{32}^{(1)}$ corresponds to the stationary solution of the
system Eq.~(\ref{rho_aver_eq}) in the linear approximation and it is
responsible for the linear susceptibility of the medium. The
nonlinear corrections $\bar\rho_{32}^{(2)}$ and
$\bar\rho_{32}^{(3)}$ determine resonant nonlinear atomic
 susceptibility. The calculation shows
that $\bar\rho_{32}^{(2)}$ it is negligible comparing with the Kerr
type nonlinearity $\bar\rho_{32}^{(3)}$, Hence, we neglect the
second order correction $\bar\rho_{32}^{(2)}$ in the paper.

We study the dynamics of $|2\rangle$ and  $|3\rangle$ levels of
sodium atoms interacting with the probe field. We characterize the
transition $|2\rangle \rightarrow |3\rangle$ in terms of the
effective coupling constant associated with the dipole matrix
element $\bar\rho_{32}$. Considering all atoms being initially in
the state $|2\rangle$, i.e. $\bar\rho_{22}=1,
\bar\rho_{11}=\bar\rho_{33}=0$ we find from Eq.~(\ref{rho_aver_eq})
\begin{eqnarray}
\label{rho32} \bar\rho_{32}^{(1)}=\frac{1}\Gamma,\;
\bar\rho_{32}^{(3)}=\frac{i}\Gamma
\frac{\Gamma^*-\Gamma}{2|\Gamma|^2}
\left({\frac1{2\gamma_{opt}}+\frac1{\gamma_{mag}}}\right),\;
\end{eqnarray}
where
\begin{eqnarray}
\label{gamma} \Gamma=\Delta-i2\gamma_{opt}+
\frac{|g_1|^2}{i\gamma_{mag}-\Delta},\\
\gamma_{opt}=\frac{\gamma_{32}+\gamma_{31}}2,\;\gamma_{mag}=\gamma_{12}.
\end{eqnarray}
A total polarization vector of the BEC coupled to the probe
electromagnetic field is given in the form
$\vec{\mathbf{P}}=\vec{\mathbf{P}}^{(l)}+\vec{\mathbf{P}}^{(nl)}$
\cite{Agrawal:2001}. Here $\vec{\mathbf{P}}^{(l)}=\epsilon_0
\hat{\chi}^{(1)} \cdot \vec{\mathbf{E}}$ describes the linear
contribution, and
\begin{eqnarray}
\label{chi} \vec{\mathbf{P}}^{(nl)}=\epsilon_0
\left({\hat{\chi}^{(2)}: \vec{\mathbf{E}} \vec{\mathbf{E}}+
\hat{\chi}^{(3)}: \vec{\mathbf{E}} \vec{\mathbf{E}}
\vec{\mathbf{E}}+ \ldots}\right)
\end{eqnarray}
is the nonlinear response on the external fields. The hat
 denotes tensors. Using well-known relation for the polarization
 induced in a resonance medium $\vec{\mathbf{P}}=\frac{\cal N}V\mu_{32}
 \bar\rho_{32}$ \cite{Wolf:2002} and Eq.~(\ref{rho32}) we find
first and third order  nonlinear susceptibilities of the
Bose-Einstein condensate:
\begin{eqnarray}
\label{chi1_3} &\chi^{(1)}=\frac{\cal
N}V\frac{|\mu_{32}|^2}{\epsilon_0\Gamma},\nonumber\\
&\chi^{(3)}=i\frac{2{\cal N}}{3V} \frac{|\mu_{32}|^4}{\epsilon_0}
\frac{\Gamma^*-\Gamma}{\Gamma|\Gamma|^2}
\left({\frac1{2\gamma_{opt}}+\frac1{\gamma_{mag}}}\right).
\end{eqnarray}
Here ${\cal N}$ is a number of atoms in BEC, $V$ is the volume, and
 ${\cal N}/V$ is the density of the condensate. Earlier, a
similar form for the linear susceptibility $\chi^{(1)}$ of BEC in
the EIT regime was obtained in the paper \cite{Hau:1999}. The
nonlinear part $\chi^{(3)}$ of Kerr type was studied in the regime
of giant nonlinearities induced in a cyclic process of
$\Lambda$-type interaction between optical fields and "hot"
$^{87}\mathrm{Rb}$ atoms in atomic cells inside an optical ring
cavity \cite{Wang:2001}.

The permittivity of Bose gas corresponding to the probe field and
including linear and nonlinear terms, reads \cite{Agrawal:2001}
\begin{eqnarray}
\label{eps} \epsilon_p=1+\chi^{(1)}+\frac34 \chi^{(3)}|A_p|^2.
\end{eqnarray}
Hence, using the relation $\epsilon_p=\left({n_p+i\frac{\eta_p
c}{2\omega_p}}\right)^2$  and Eq.~(\ref{chi1_3}) we find the
refraction index $n_p$ and the absorption coefficient $\eta_p$ in
first order with respect to the intensity of the probe field
\begin{eqnarray}
\label{n_p} &n_p=n_p^{(0)}+n_p^{(2)} |A_p|^2,\;
n_p^{(0)}=1+\frac12 \mathrm{Re}(\chi^{(1)}),\nonumber\\
&n_p^{(2)}=\frac38 \mathrm{Re}(\chi^{(3)});\\
&\eta_p=\eta_p^{(0)}+\eta_p^{(2)} |A_p|^2,\;
\eta_p^{(0)}=\frac{\omega_p}c  \mathrm{Im}(\chi^{(1)}),\nonumber\\
&\eta_p^{(2)}=\frac{3\omega_p}{4c}  \mathrm{Im}(\chi^{(3)}).
\end{eqnarray}

In the work, the Bose-Einstein condensate is described by the
$\Lambda$-scheme in near to resonance regime. The concentration of
sodium atoms in the condensate ${\cal N}/V=3.3\cdot 10^{12}
\mathrm{cm^{-3}}$ is taken from \cite{Hau:1999} and the coupling
field intensity $I_c=55 \mathrm{\frac{mW}{cm^2}}$. Taking the dipole
matrix element $|\mu_{32}|=22\cdot10^{-30} \mathrm{C\cdot m}$ from
\cite{Babiker:1998} and making use of the definition
$A_c=\sqrt{\frac{2I_c}{c\epsilon_0}}$ \cite{Agrawal:2001} we
calculate the coupling constant ${g_1}/(2\pi)=21.4 \mathrm{MHz}$.
The probe pulse  is assumed to have a time-length of $1 \mathrm{\mu
s}$, the laser waist inside the BEC is $d=3.7 \mathrm{\mu m}$. The
intensity of the probe pulse $I_p=80 \mathrm{\frac{\mu W}{cm^2}}$
corresponds to 25 photons in average.

Due to atomic coherence in BEC, in absence of the Doppler broadening
the decay rates $\gamma_{31}$ and $\gamma_{32}$ of the level
$|3\rangle$ can be estimated by natural (spontaneous) width of
transitions from the upper levels of sodium atoms. Taken from the
paper \cite{Babiker:1998} the lifetime on the upper level is
$T_{rel}=16.3 \mathrm{ns}$, so we assume $\gamma_{32} /
2\pi=\gamma_{31} / 2\pi=5 \mathrm{MHz}$. The decay rates of
transitions between the hyperfine levels $|1\rangle$ and $|2\rangle$
is ${\gamma_{12}}/({2\pi})=38 \mathrm{KHz}$ according to
\cite{Wolf:2002}.

\begin{figure}
\includegraphics[width=80mm]{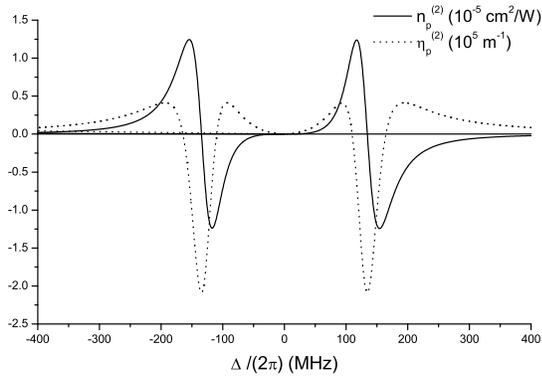}
\caption{\label{fig:spec2} The refraction index $n_p^{(2)}$ and the
absorption coefficient $\eta_p^{(2)}$ of $ ^{23} \mathrm{Na}$ BEC as
functions of the detuning $\frac\Delta{2\pi}$.}
\end{figure}

In Fig.~\ref{fig:spec2} we plot a typical frequency dependence of
the nonlinear optical refraction index and absorption coefficient
as functions of detuning $\Delta$. Giant nonlinear refraction
index formed in the condensate with appropriately chosen detuning
of the probe pulse is exploited in the paper to demonstrate the
possibility of generating subpoissonian statistics of photons on
very short time-scales. It is also worth noticing that the
realization of a negative $n_p^{(2)}$ is similar to the effects
observed in \cite{Wang:2001}, and it may have some important
physical applications.

On the other hand, the alternation of regions with positive and
negative absorption coefficient corresponds to the regimes of
effective nonlinear reduction or amplification of the probe field
intensity in the BEC. In both cases, the energy is transformed
between the coupling field and the probe pulse. The case of zero
absorption $\eta_p=0$ may be characterized as a nonlinear EIT.
Notice that the regime of complete absence of optical losses in the
system falls at the region of very large values of $n_p$. It opens
even more extensive perspectives to generate nonclassical light in
the system. In generally, varying the intensity of the coupling and
probe light, and the detuning $\Delta$ we can control the parameter
$\Gamma$ in Eq.~(\ref{chi1_3}) and change linear and nonlinear
relative impact on the probe pulse in the condensate.

\begin{figure}
\includegraphics[width=80mm]{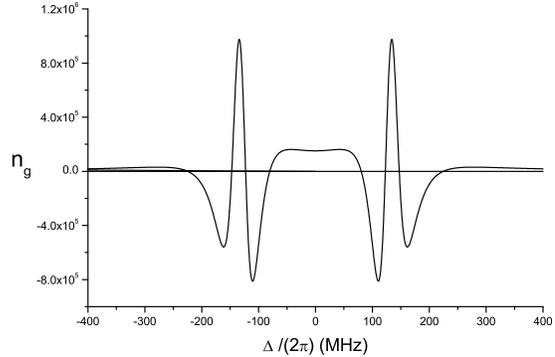}
\caption{\label{fig:spec3} The group refraction index $n_g$ as a
function of the detuning $\frac\Delta{2\pi}$.}
\end{figure}

To understand the effect of "slow" light in the system we study a
frequency dependence of some characteristics of the probe pulse
envelope. In Fig.~\ref{fig:spec3} we plotted the group refraction
index $n_g=n_p+\omega_p \frac{dn_p}{d\omega_p}$. A point $\Delta=0$
 of the exact resonance corresponds to the EIT regime \cite{Harris:1990}
characterized by low losses and large $n_g$. Under chosen resonance
conditions, the group velocity $v_g=c/n_g$ is $2000 m/s$ while the
losses are $\eta_p=242 m^{-1}$. So, the regime corresponds to the
effect of "slow" light. An interesting point in Fig.~\ref{fig:spec3}
is a zero of the group refraction index $n_g$ that indicates an
uncertainty of the group velocity $v_g$ of the probe pulse. The
uncertainty depicts a possibility to observe superluminal
velocities. This effect together with the negative group refraction
index is explained in the literature \cite{Garrett:1970,
Mitchell:1997} by the presence of plane waves of different
frequencies, which appear in the medium long before the pulse peak
enters into it. When the group velocity is negative ($n_g<0$) the
pulse generates two anti-propagating pulses in the BEC.  One of them
propagates backward (negative velocity) and prevents the peak from
traveling forward in the medium.

In the following sections we concentrate on some effects in dynamics
of the weak probe pulse. These effects are induced in the system of
highly correlated three-level particles by relatively strong
classical coupling field. We show how the later can control some
quantum properties of the probe pulse.

\section{The effective two-level quantum model} \label{sec:part2}

We are interested in quantum properties of the probe field and
define the field in terms of creation and annihilation operators
$a^\dagger, a$. Considering only two levels $|2\rangle$ and
$|3\rangle$ of interest for the dynamics of the probe field we work
in the EIT limit, i.e. most of the atoms is concentrated on the
lower level $|2\rangle$. We assume that the classical field is
strong and describe its influence in terms of effective coupling
constant for the probe field. The effective two-level Hamiltonian in
dipole and rotating-waves approximations reads
\begin{eqnarray}
\label{Hamilt_q} &H=\omega_p(a^\dagger a +S_3+\frac{\cal
N}2)+\Delta S_3 +k_1(aS_+
+ a^\dagger S_-) \nonumber \\
&+k_2(a^\dagger a a^\dagger S_- +a a^\dagger a S_+).
\end{eqnarray}
The operators $S_\pm, S_3$ describe total dipole momentum
corresponding to the transitions $|3\rangle\leftrightarrow|2\rangle$
for the atoms in BEC. The first and the second terms give the free
energy of the probe field and atoms. The third term describes linear
contribution into the interaction between the field and two-level
particles in dipole approximation. It has the typical form of so
called the Tavis-Cummings \cite{Tavis:1968} or Dicke model
\cite{Dicke:1954}. The last term in Eq.~(\ref{Hamilt_q}) describes
nonlinear processes due to the presence of strong classical coupling
field, and it depends on the intensity  of the probe pulse. Both
coupling constants are defined below.

An interaction between the BEC and the probe field is determined by
the effective matrix element of atomic dipole momentum between
$|2\rangle$ and $|3\rangle$ levels. Due to induced nonlinearity the
matrix element depends on the intensity of the probe field described
by the operator $a^\dagger a$ and on the intensity of the coupling
field given by a c-number parameter. We define overall coupling
constant in the dipole approximation as follows
\begin{equation}\label{k_gen}
    k_0
    \frac{\bar\rho_{32}(g_1,\Delta,g_2)}
    {\bar\rho_{32}^{(1)}(g_1=0,\Delta) g_2},
\end{equation}
where $\bar\rho_{32}$ is given in Eq.~(\ref{rho_nonlin}), and
$k_0=\mu_{32}\sqrt{\frac\omega{2\hbar\varepsilon_0 V}}$ is the
single-photon Rabi-frequency in the Dicke model\cite{Andre:2002}.
The form-factor in Eq.(\ref{k_gen}) satisfies the condition that at
zero coupling field it approaches unity. This behavior of the
form-factor results from the fact that when $g_1\approx0$ the
induced nonlinearity for the probe field should vanish.

From the definition Eq.(\ref{k_gen}) and the expansion of
$\bar\rho_{32}$ in Eqs.~(\ref{rho_nonlin}), (\ref{rho32}) we find
the coupling constants $k_1$ and $k_2$
\begin{eqnarray}
\label{effect_coupl} k_1=k_0 L_l,\;k_2=k_0^3 L_{nl},
\end{eqnarray}
The parameters $L_{l(nl)}$ denote linear and nonlinear
form-factors (compare with \cite{Icsevgi:1969}), correspondingly
\begin{eqnarray}
\label{form_fact} L_l=\frac{\bar\rho_{32}^{(1)}(g_1,\Delta)}
    {\bar\rho_{32}^{(1)}(g_1=0,\Delta)},\;
L_{nl}=\frac{\bar\rho_{32}^{(3)}(g_1,\Delta)}
    {\bar\rho_{32}^{(1)}(g_1=0,\Delta)}.
\end{eqnarray}
As far as the phase of the probe field is arbitrary we can choose
it in such a way that $k_1$ would be purely real with $\tilde
L_l=|L_l|$. In that case the nonlinear coupling constant would be
defined with $\tilde L_{nl}=e^{-i \arg(L_l)} L_{nl}$.

According to this formulation we are able to dynamically control
the rates of the transitions induced by the quantum probe field
$A_p$ with help of the classical field $A_c$. Changing the latter
one can reach qualitatively different regimes of the quantum
dynamics described by the effective Hamiltonian
Eq.~(\ref{Hamilt_q}).

The effective Hamiltonian can be expressed in terms of fifth-order
polynomial algebra of excitations (PAE)~\cite{pae:2003}. The
generators $M_0 ,M_\pm $ of the algebra are realized as follows
\begin{eqnarray}
\label{eq16} M_0 = \frac{a^\dagger a - S_3}{2},\; M_ + =
(1+\frac{k_2}{k_1}a^\dagger a)a^\dagger S_ ,
\end{eqnarray}
\noindent and $M_ - = (M_+)^\dagger$. These generators $M_0,M_\pm$
satisfy basic commutation relation for any PAE
\begin{equation}
\label{com1} \left[ {M_0 ,M_\pm } \right] = \pm M_\pm,
\end{equation}
and commute with the operators
\begin{equation} \label{eq17}
M = a^\dagger a + S_3 + r, \; S^2 = S_3 ^2 +
\frac{1}{2}\left( {S_ + S_ - + S_ - S_ + } \right).
\end{equation}
Hereafter we use the same notation $M$ both for the Casimir operator
and its eigenvalue, if no confusion arises. It is known that the
eigenvalues $M, r(r+1)$ of the operators in Eq.(\ref{eq17})
parameterize the PAE in question. We thus denote this PAE by
$\mathbb{M}_{M,r}$. The structure polynomial of $\mathbb{M}_{M,r}$
can be expressed in the form

\begin{eqnarray}
& p_5 \left( {M_0 } \right)=M_ + M_ - = a^\dagger a
(1+\frac{k_2}{k_1}a^\dagger a)^2 \left( {S^2 - S_3 ^2 - S_3 }
\right)\nonumber\\
& =  -\left( {M_0 + \frac{M - r}{2}} \right) \left(
{1+\frac{k_2}{k_1}M_0 +\frac{k_2}{k_1}\frac{M - r}{2} } \right)^2
\nonumber\\
&\left( {M_0 - \frac{M - 3r}{2}} \right)\left( {M_0 - \frac{M +
r+2}{2}} \right).
 \label{eq19}
\end{eqnarray}

\noindent The parameters of this structure polynomial are
\begin{eqnarray}
\label{eq20} &c_0 = - \left({\frac{k_2}{k_1}}\right)^2,\;
q_0 = -\left({\frac{k_1}{k_2}+\frac{M -r}{2}}\right),\;q_1 = -\frac{M -r}{2}\nonumber \\
&q_2 =\frac{M - r}{2} -r,\;q_3 = \frac{M - r}{2} + r + 1.
\end{eqnarray}
\noindent Notice that the root $q_0$ is a degenerate root of order
2.

Following standard procedure~\cite{pae:2003} we describe physically
interesting finite dimensional irreducible representation (irrep) of
$\mathbb{M}_{M,r}$. In our model the parameter $r$ has the meaning
of collective Dicke index (an analog of the orbital momentum) of the
system of ${\cal N}$ two-level particles. This index runs from
$\varepsilon({\cal N}) = \frac{1 - \left( { - 1} \right)^{{\cal N}
}}{4}$ to $\frac{{\cal N} }{2}$ with unit steps, while $M$ can be
any natural number including zero that follows from the definition
Eq.(\ref{eq17}). Physically interesting irrep of $\mathbb{M}_{M,r}$
is characterized by two roots of the structure polynomial where it
takes positive values in between. The roots $q_{1,2,3}$ do not
depend on the ratio $\frac{k_1}{k_2}$ and are ordered according to
the relation between $M$ and $r$. For $M\geq 2r$ $q_1<q_2\leq q_3$,
and for $M<2r$ $q_2< q_1<q_3$. Since the root $q_0$ is of order 2,
its position on the real axis doesn't influence the region where the
structure polynomial is nonnegative. Hence, if $M>2r$, the irrep is
called a remote zone and $p_5(x)$ is nonnegative in the interval
$[q_2,q_3]$, whereas if $M<2r$ the irrep is called a nearby zone and
$p_5(x)$ is nonnegative in the interval $[q_1,q_3]$. Notice that the
region $M\gg2r$ is usually called the strong-field limit and the
region $M\ll 2r$ is usually called the weak-field limit. The case
$M=2r$ is of special kind and the corresponding irrep is called the
boundary zone.

Depending on the value of the ratio $\frac{k_1}{k_2}$ we might have
two different situations. If $q_0$ doesn't belong to the interval
where $p_5(x)\geq 0$, the polynomial is approximated by the
parabolic function relatively well that correspond to the method
described in our previous paper~\cite{pae:2003}. In case $q_0$
belongs to the interval of positiveness of $p_5(x)$ we must
introduce some changes to the approach. But for the physical system
considered here the ratio takes large negative values and the total
number of excitations $M$ may be considered being smaller than the
ratio, i.e. $M<|\frac{k_1}{k_2}|-1$. Therefore, $q_0$ is always
larger than $q_3$ and we can use the algebraic approach developed
for conventional Tavis-Cummings model.

In the physical situation being studied here, the remote and nearby
zones are bounded by non-degenerate roots of $p_5(x)$ and the
corresponding physical irrep of the fifth-order PAE
$\mathbb{M}_{M,r}$ is isomorphic to physical irrep of second-order
PAE, denoted in the paper as $\mathbb{S}_{\tilde r}$. We use here
$\tilde{r}$ to distinguish it from collective Dicke index $r$
describing the two-level system. We solve the eigenvalue problem of
the Hamiltonian Eq.(\ref{Hamilt_q}) in terms of simpler algebra
$\mathbb{S}_{\tilde r}$ making use of the isomorphism. The idea is
straight forward. The generators $M_\pm, M_0$ are realized in terms
of the generators $\tilde S_\pm, \tilde S_3$ of $\mathbb{S}_{\tilde
r}$ according to the isomorphism and substituted into the
Hamiltonian. The interaction part is expanded into perturbation
series of power of the operator $\tilde S_3$ and the series are
diagonalized by consecutive unitary transformations.

To begin with we consider the  transformation of
$\mathbb{M}_{M,r}$ to $\mathbb{S}_{\tilde r}$ for the case of
remote zones. The dimension of a remote zone is $2r + 1$. Hence,
the algebra $\mathbb{S}_{\tilde r}$ is characterized by
$\tilde{r}=r$. The finite dimensional irrep of $\mathbb{S}_{\tilde
r}$ is isomorphic to the corresponding irreducible representation
of su(2) algebra. The corresponding transformation from the
generators of $\mathbb{M}_{M,r}$ to the generators of
$\mathbb{S}_{\tilde r}$ is defined as follows (see
~\cite{pae:2003})
\begin{eqnarray}
 & M_0 = \frac{M - r}{2} - \tilde {S}_3,\\
&M_ + =\tilde{S}_ - \sqrt {M - r +1 - \tilde
{S}_3}\left({1+\frac{k_2}{k_1}(M-r+1-\tilde S_3)}\right).\nonumber
\label{eq28}
\end{eqnarray}

\noindent Spectrum $\{\tilde {m}\}$ of the operator $\tilde{S}_3$
belongs to the interval $[-r,r]$ so the argument of the square
root function in Eq.(\ref{eq28}) has positive valued spectrum in
the remote zones $(M>2r)$. Expanding the root function with
respect to $(\tilde S_3-\frac12)$ we obtain the perturbation
series with smallness parameter $\alpha=\frac1{M-r+\frac12}$.

It is worth to notice the connection between new operators $\tilde
S_\pm, \tilde S_3$ and the physical operators of the model. From
Eqs.(\ref{eq16}),(\ref{eq17}), and (\ref{eq28}) it follows that in
remote zones
\begin{equation}
\label{eq29a} \tilde {S}_3 = S_3,\; \tilde {S}_+ =
\frac1{\sqrt{a^\dagger a+1}}\; a\cdot S_+, \; \tilde {S}_- = \left
({\tilde {S}_+}\right )^\dagger .
\end{equation}
Notice that the subspaces corresponding to the remote zones do not
contain the vacuum state of the field. It is also worth mentioning
that the matrix representation of the operator
$\frac1{\sqrt{a^\dagger a+1}}a$ is $\delta_{n,n+1}$ in any remote
zone. This operator has been considered before in terms of phase
operator \cite{Chumakov:1994, Saavedra:1998, Delgado:2001}.

We turn now to the nearby zones $M < 2r$. Notice that the
dimension of nearby zones is $q_3-q_1=M+1$, and therefore
$\tilde{r}=\frac{M}{2}$. Hence, we obtain the following
realization
\begin{eqnarray}
& M_0 = \frac{r}{2} - \tilde{S} _3, \, M_- = (M_+)^\dagger  ,\nonumber\\
&M_ + = \tilde{S} _ - \,\sqrt {\frac{4r - M}{2} +1 - \tilde{S}
_3}\left({1+\frac{k_2}{k_1}(\frac M2+1-\tilde S_3)}\right).\quad
\label{eq34}
\end{eqnarray}
 Since all the eigenvalues of the operator $\tilde{S} _3 $ belong
 to the interval $[ - \tilde{r},\tilde{r}]$, the argument of the square root function
does not have zero eigenvalues in the nearby zones. The
realization of $\mathbb{S}_{\tilde r}$ through spin and boson
variables is then given by
\begin{eqnarray}
\tilde{S} _3 = \frac M2 - a^\dagger a,\; \tilde{S} _ +
=\frac1{\sqrt{r+1-S_3}} S_+ \, a. \label{eq34a}
\end{eqnarray}
\noindent Notice that the nearby zones do not contain the
eigenvector $|r,r\rangle$  of $S_3$.

\section{Diagonalization procedure}
\label{sec:part4}

We start with the representation of the Hamiltonian
Eq.(\ref{Hamilt_q}) in the form of series of the operator
$\alpha(\tilde S_3 -\frac12)$, where the constant $\alpha$ is a
smallness parameter being specified below. As we already
mentioned, to construct the series we expand the square root
function in the realizations Eq.(\ref{eq28}) and Eq.(\ref{eq34})
depending on the zone under consideration. The smallness parameter
has the form
\begin{equation}
\label{alpha} \alpha \equiv \left\{ {{\begin{array}{c}
  \frac1{M - r + \frac{1}{2}}, \;M\geq 2r \\ \\
  \frac2{4r - M + 1}, \; M<2r \\
\end{array}}}\right. .
\end{equation}
\noindent An accuracy necessary to observe all the interesting
effects described in the paper, is provided by first three terms in
the expansion of the effective Hamiltonian  Eq.(\ref{Hamilt_q}). Up
to second order with respect to the smallness parameter $\alpha$ it
reads
\begin{eqnarray}
&H\approx\omega_p(M+\frac{\cal N}2-r)+\Delta (\tilde S_3 +\tilde r - r)+ \nonumber\\
&k\left[{\begin{array}{c}
  \frac{ \tilde S_+ + \tilde S_-}2 - \frac{\beta_1}4\left(
{\left( {\tilde S_3  - \frac12} \right)
 \tilde S_ +   + \tilde S_ -  \left( {\tilde S_3 - \frac12}
 \right)} \right) \\ - \frac{\beta_2}2
  \left( {\left( {\tilde S_3  - \frac12} \right)^2 \,
  \tilde S_ +   + \tilde S_ -  \,\left( {\tilde S_3  - \frac12}
   \right)^2 } \right)\end{array}}\right].\quad\label{eq31}
\end{eqnarray}
Here we used a simple relation $S_3=\tilde S_3+\tilde r-r$. To
prove it one should utilize the definition of $\tilde r$ in the
remote or nearby zones given in the previous section, and the
relations Eqs.(\ref{eq29a}) and (\ref{eq34a}). The parameters in
Eq.(\ref{eq31}) are defined as follows:
\begin{eqnarray}
\label{params1} \gamma = \left\{ {{\begin{array}{c}
  1+\frac{k_2}{k_1}(M - r + \frac12), \,M\geq 2r \\ \\
  1+\frac{k_2}{k_1}\frac{M + 1}{2}, \quad M<2r \\
\end{array}}}\right.
\end{eqnarray}

\begin{eqnarray}
&k = k_1 \gamma \frac2{\sqrt\alpha},\nonumber\\
&\beta_1=\alpha(1+\frac{k_2}{k_1}\frac2{\gamma\alpha}),\,
\beta_2=\alpha^2(\frac18-\frac{k_2}{k_1}\frac1{2\gamma\alpha}).\quad
\label{params2}
\end{eqnarray}
Above, considering the irreducible representations of the
fifth-order algebra $\mathbb{M}_{M,r}$ we noticed a limitation on
the value $M$ provided that the algebraic approach for the
Tavis-Cummings model is applicable. In other words, relatively
accurate approximation of the structure polynomial $p_5(x)$ by the
parabolic polynomial of $\mathbb{S}_{\tilde r}$ in the interval
between two roots where $p_5(x)$ takes positive values is
achievable if $q_0$ doesn't belong to the interval. Now we are
able to define the limitation of the algebraic approach more
rigorously. We find from Eqs.(\ref{params1}),(\ref{params2}) that
$\gamma\sim O(1)$ and $\beta_n\sim \alpha^n$ if
\begin{equation}
\label{valid} M+1<|\frac{k_1}{k_2}|.
\end{equation}

The Hamiltonian Eq.(\ref{eq31}) may be rearranged into the
following form
\begin{eqnarray}
&H^{(2)}= C_0+\Delta \tilde S_3 + \nonumber\\
&k\left[{\tilde S_x - \frac{\beta_1}4\left({ \tilde S_3 \tilde S_x
+ \tilde S_x  \tilde S_3} \right) - \beta_2\left({\tilde S_3\tilde
S_x \tilde S_3+\frac14 \tilde S_x}\right)}\right],\quad
\label{Hnew}
\end{eqnarray}
where $\tilde S_x=\frac{\tilde S_+ +\tilde S_-}2$ according to the
su(2) algebra notations, and $C_0=\omega_p(M+\frac{\cal
N}2-r)+\Delta (\tilde r - r)$ is a constant in each irreducible
representation because M is the Casimir operator.

The first three terms in Eq.(\ref{Hnew}) are linear with respect
to the generators of $\mathbb{S}_{\tilde r}$ and can be
diagonalized by well-known su(2) unitary transformation $U_0 =
e^{(i\psi_0 \tilde S_y)}$ corresponding to rotation of the
quasi-spin vector $(\tilde S_x,\tilde S_y, \tilde S_3)$ about
y-axis, where $\tilde S_y=\frac{\tilde S_+ - S_-}{2i}$. The angle
$\psi_0$ is found from the relations
\begin{equation}
\cos(\psi_0) = \frac\Delta{\Omega_R},\;
\sin(\psi_0)=\frac{k}{\Omega_R},\; \Omega _R
=\sqrt{\Delta^2+k^2}\, , \label{U0}
\end{equation}
where we introduced a notion of nonlinear quantum Rabi frequency
$\Omega _R$. Hence, after the transformation $U_0 H^{(2)}
U_0^{-1}$ a zero order contribution into Eq.(\ref{Hnew}) with
respect to $\alpha$ reads $(C_0+\Omega_R \tilde S_3)$ that
justifies the name for the constant $\Omega_R$. It describes the
frequency of rotation of the quasi-spin $\tilde r$ of the
atom-field quantum system.

Applying two additional unitary transformations $U_1, U_2$
discussed in Appendix \ref{app:diag}, we diagonalize the
Hamiltonian up to second order with respect to $\alpha$. The
diagonalized operator takes the form:

\begin{widetext}
\begin{eqnarray}
&H^{(2)}_{diag}(\tilde S_3)=(U_2 U_1 U_0) H^{(2)}(U_2 U_1
U_0)^{-1}= C_0+\Omega_R\tilde S_3- \frac{\beta_1}4
\frac{k^2\Delta}{\Omega_R^2}\left({3\tilde S_3^{\,2}- \tilde
r(\tilde r+1)}\right) +\nonumber \\
& \left({\frac{\beta_1}4}\right)^2 \frac{k^2}{\Omega_R}\tilde
S_3\left[{\frac{4\Delta^4 -
 9 \Delta^2k^2+4k^4}{\Omega_R^4}\tilde S_3^{\,2}-\frac{2\Delta^4 -
 5\Delta^2k^2+2k^4}{\Omega_R^4} \tilde r(\tilde r+1)+\frac12 \frac{\Delta^4
 + \Delta^2k^2+k^4}{\Omega_R^4}}\right] -\nonumber \\
&\frac{\beta_2}2 \frac{k^2} {\Omega_R} \tilde
S_3\left[{\frac{4\Delta^2 - k^2} {\Omega_R^2} \tilde
S_3^{\,2}-\frac{2\Delta^2 - k^2} {\Omega_R^2} \tilde r(\tilde
r+1)+\frac12 \frac{\Delta^2 -k^2} {\Omega_R^2}}\right]
+o(\alpha^2) . \label{eq36}
\end{eqnarray}
\end{widetext}

\begin{figure}
\includegraphics[width=80mm]{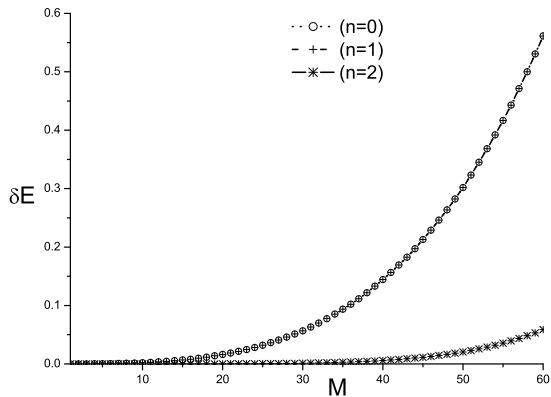}
\caption{\label{fig:spec4} Relative error, $\delta
E(M)=\max(\frac{E(M)-E^{(n)}(M)}{\Delta E(M)})$, for three different
orders of accuracy ($n=0,1,2$). Here, $E(M)$ is the exact spectrum
of the Hamiltonian Eq.(\ref{Hamilt_q}) calculated numerically in the
irrep specified by given M and r values, $\Delta E(M)$ is an average
splitting between the spectrum levels, and $E^{(n)}(M)$ is the
spectrum calculated by Eq.(\ref{eq36}) in n-th order with respect to
$\alpha$. For the condensate, we choose ${\cal N}=1000$, hence $r
={\cal N}/2$. For the field, the parameters
$\Delta/\omega_p=2.4\mathrm{E\!-\!8}$,
$k_1/\omega_p=3.04\mathrm{E\!-\!7}$,
$k_2/\omega_p=-3.01\mathrm{E\!-\!9}$ are taken according to the
consideration in the previous section.}
\end{figure}

In Fig.~\ref{fig:spec4} we compare the second order solution
Eq.(\ref{eq36}) with the exact numerical diagonalization of $H$
Eq.(\ref{Hamilt_q}) and find that the approximation is very accurate
and doesn't significantly depend on the number $\cal N$ of the
particles in the condensate. Basically, as higher the ratio
$\frac{|M-2r|}{2r}$ is as better the analytical solution
Eq.~(\ref{eq36}) becomes. The first order correction to the spectrum
(of order $\beta_1$ in Eq.~(\ref{eq36})) is proportional to the
detuning $\Delta$ and vanishes at the point of exact resonance.
Therefore, the relative error decreases significantly only if the
second order correction is taken into account. One can understand
from Fig.~\ref{fig:spec4} that at the points where zero and first
order errors increase to 50\% our second order correction provides
accuracy even higher than 95\%. The plot depicts typical behavior of
the second order solution in the nearby zones. According to the
condition Eq.~(\ref{valid}) the approximation diverges when we
approach too close to the point $|k_1/k_2|-1=100$. Therefore, we
restrict our initial state of the system to the maximum of 60
excitations. It is justified in the next section from physical point
of view.

\section{Subpoissonian distribution in photon statistics} \label{sec:part5}
To study photon statistics we have to construct the time evolution
of corresponding field operators. It is well-known that the
simplest characteristic of subpoissonian nature of the photon
distribution is the Fano-Mandel parameter Q(t) defined as follows
\begin{equation}
\label{Mandel} Q(t) = \frac{\langle (a^\dagger a)^2\rangle -
\langle a^\dagger a\rangle^2} {\langle a^\dagger a\rangle} - 1 .
\end{equation}
Our unitary transformation approach allows to represent the
averages of field operators in Eq.(\ref{Mandel}) in the following
convenient form:
\begin{eqnarray}
\label{t_evol} &\langle A(t)\rangle = \langle\Phi_0|e^{i Ht} A
e^{-i Ht}|\Phi_0\rangle\approx \nonumber\\
&\langle\Phi_0| U_0^{-1} \left({ e^{i H_{diag}^{(2)}t}U_0 A
U_0^{-1} e^{-i H_{diag}^{(2)}t}}\right)U_0|\Phi_0\rangle,
\end{eqnarray}
where $|\Phi_0\rangle$ is an initial state of the atom-field
system and $A$ is an arbitrary operator. In Eq.~(\ref{t_evol}) we
left only zero order terms with respect to $\alpha$, i.e.
$U_{1,2}\rightarrow 1$, but the diagonalized Hamiltonian
$H_{diag}^{(2)}(\tilde S_3)$ is kept up to second-order terms
because the time intervals of interest may be relatively large.
Applying the relation $a^\dagger a=M-\tilde S_3-\tilde r$ it is
straight forward to find the time dependence of $A=a^\dagger a$
from Eq.(\ref{t_evol}). The result reads

\begin{eqnarray}
\label{N_phot} & e^{i H_{diag}^{(2)}t}U_0\, a^\dagger a\, U_0^{-1}
e^{-i H_{diag}^{(2)}t} = M-\tilde r -\frac\Delta{\Omega_R}\tilde
S_3+\nonumber \\
&\frac k{2\Omega_R} \left({\tilde S_+ e^{it
\left[{H_{diag}^{(2)}(\tilde S_3+1)- H_{diag}^{(2)}(\tilde
S_3)}\right]} + h.c.}\right).
\end{eqnarray}
Taking the square of the expression Eq.(\ref{N_phot}) one easily
finds corresponding formulae for $A=(a^\dagger a)^2$. To calculate
the parameter Q(t) we have to average these operators over unitary
transformed initial state, i.e. over $U_0|\Phi_0\rangle$. The
transformation $U_0$ is studied in detail in the theory of the su(2)
algebra. If we succeed to represent the initial atom-field state
$|\Phi_0\rangle$ in the basis of eigenstates $|\tilde m, \tilde
r\rangle$ of the operator $\tilde S_3$ then the corresponding
amplitudes $B_{\tilde m, \tilde m1}$ in the expansion $|\tilde m,
\tilde r\rangle_{dr}\equiv U_0|\tilde m, \tilde
r\rangle=\sum_{\tilde m1=-\tilde r}^{\tilde r} B_{\tilde m, \tilde
m1}|\tilde m1, \tilde r\rangle$ can be taken from a textbook. The
state $|\tilde m, \tilde r\rangle_{dr}$ is usually called
generalized atom-field dressed state.

In the paper we choose a typical initial state of the system, which
seems natural for the experiments with sodium BEC. The probe field
is prepared in a coherent state with average number of photons
$n_0=25$. The atoms are prepared in completely symmetrized unexcited
eigenstate of the operator $S_3$, i.e. $|m_0,r={\cal N}/2\rangle$,
with zero number of excitations ($m_0+\frac{\cal N}2=0$). Hence,
\begin{eqnarray}
\label{init_st} &|\Phi_0\rangle=|-\frac{\cal N}2, \frac{\cal
N}2\rangle\otimes \sum\limits_{n=0}^\infty
\sqrt{e^{-n_0}\frac{n_0^n}{n!}}\,|n\rangle\nonumber\\
&=\sum\limits_{n=0}^\infty
\sqrt{e^{-n_0}\frac{n_0^n}{n!}}\,|\tilde m=-\frac n2, \tilde
r=\frac n2\rangle.
\end{eqnarray}
The last equality follows from the relations Eqs.~(\ref{eq17}) and
(\ref{eq34a}), because for completely unexcited atoms the total
number of excitations $M$ is equal to the number of photons $n$.
 Having represented the initial state in the basis of the
eigenvectors of the operator $\tilde S_3$ we can immediumtely
calculate the Fano-Mandel parameter $Q$. Below in
Fig.~\ref{fig:spec5} we plot the evolution of Q(t) and the average
number of photons for such initial state. Regions where the
parameter $Q(t)$ takes negative values correspond to subpoissonian
distribution in statistics of photons in the probe pulse. First, we
notice that the minimum of $Q(t)$ is reached after the second Rabi
oscillation, which is described by corresponding Rabi frequency
$\Omega_R$ given in Eq.~(\ref{U0}). The frequency depends on the
number of excitations $M$. Being calculated for the average initial
number of excitations $M=n_0$ it gives the period of Rabi
oscillation $T_R=2\pi/\Omega_R=0.12 ns$, which is confirmed by
Fig.~\ref{fig:spec5} (see the dynamics of the average number of
photons). The time interval where the probe pulse obtains maximum
squeezing in fluctuations of the number of photons at, is
approximately 0.5ns that is much smaller than the relaxation times
in BEC ($T_{rel}=16.3 ns$). This time scale satisfies the adiabatic
condition imposed at the beginning. The dispersion of photons in the
initial coherent state is equal to the average number of photons,
i.e. to $n_0=25$. It means that the probability to find more than
$2n_0=50$ photons is negligible and the restriction from above on
the number of excitations $M<|k_1/k_2|\approx100$ in the system is
fulfilled as well.

\begin{figure}
\includegraphics[width=80mm]{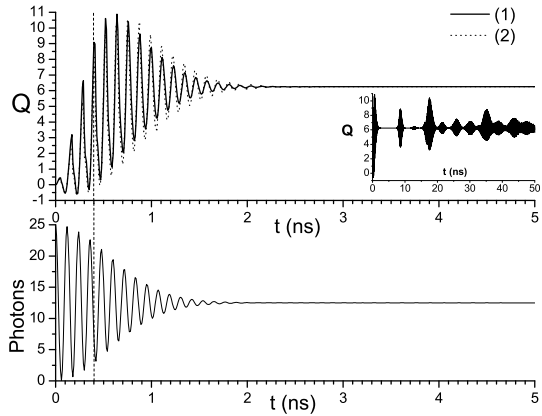}
\caption{\label{fig:spec5} The Fano-Mandel parameter Q(t) and the
average number of photons for initially unexcited sodium BEC and a
coherent probe field with $n_0=25$ average number of photons. All
the parameters are the same as in Fig.~\ref{fig:spec4}. The solid
line (1) on the plot of $Q(t)$ is calculated in second order of
$\alpha$ according to Eq.~(\ref{t_evol}). The dotted line (2) is
calculated by Eqs.~(\ref{photon_0order}) and
(\ref{photon2_0order}).}
\end{figure}
We can notice in Fig.~\ref{fig:spec5} that the maximum squeezing
is relatively large, $Q\approx-0.6$. It is observed after two
cycles of almost complete absorption and re-emission of the
photons by sodium BEC that is represented by the plot of the
average number of photons in the probe pulse in
Fig.~\ref{fig:spec5}. The quantum effect is provided by the
correlation in the atomic system, which is transferred to the
field after two complete Rabi cycles of the collective atom-field
quasi-spin $\tilde S_3$.

The inline plot of the parameter $Q(t)$ at large time scale shows
that the effect has very short life-time and only appears at the
beginning. The regular collapses and revivals are observed due to
the interference between Rabi oscillations with different
frequencies. The effect is well-known in the one-atom case, and the
collapse and revival times were estimated in the single-particle
model\cite{Eberly:1980}. We will show now that due to the nonlinear
interaction between sodium atoms and the probe pulse the maximum
squeezing of photon statistics is achieved much faster than in the
regular Dicke model. The fact plays crucial role if one takes into
account different mechanisms of the decoherence and relaxation in
the system. For instance, the adiabatic approach developed in the
paper would break down without the nonlinear terms in the
Hamiltonian.

Since we are interested in a short time-scale dynamics, we can
neglect higher order corrections to the spectrum of the Hamiltonian
Eq.~(\ref{eq36}) keeping only zero order terms in
Eq.~(\ref{t_evol}). Then, we find
\begin{widetext}
\begin{eqnarray}
\label{photon_0order} &\langle a^\dagger
a(t)\rangle=\langle\Phi_0|(M-\tilde r)(1-\cos(\Omega_R
t))\sin(\psi_0)^2|\Phi_0\rangle+\langle\Phi_0|a^\dagger
a\left({\cos(\psi_0)^2+\sin(\psi_0)^2 \cos(\Omega_R
t)}\right)|\Phi_0\rangle,
\\
\label{photon2_0order} &\langle a^\dagger
a(t)^2\rangle=\langle\Phi_0|(a^\dagger
a)^2\;\left[{\cos(\psi_0)^4+ \cos(\psi_0)^2 \sin(\psi_0)^2 (3
\cos(\Omega_R t)-1)+\frac14 \sin(\psi_0)^4 (1+3\cos(\Omega_R
t))}\right]|\Phi_0\rangle\nonumber \\
&+\langle\Phi_0|a^\dagger a\;\left[{(M-\tilde r) \sin(\psi_0)^2
\sin(\frac{\Omega_R}2 t)^2
\left({5+3\cos(2\psi_0)+ 6\sin(\psi_0)^2 \cos(\Omega_R t)}\right)}\right]|\Phi_0\rangle \\
&+\langle\Phi_0|\left[{\frac12\sin(\psi_0)^2 \sin(\frac{\Omega_R}2
t)^2 \left({3\tilde r(\tilde r+1)-(M-\tilde r)^2+  (\tilde
r(\tilde r+1)-3(M-\tilde r)^2)(\cos(2\psi_0)+2 \cos(\Omega_R
t)\sin(\psi_0)^2)}\right)}\right]|\Phi_0\rangle.\nonumber
\end{eqnarray}
\end{widetext}
Notice, that $M, \cos(\psi_0), \sin(\psi_0), \tilde r, \Omega_R$ are
not c-numbers but Casimir operators, so they have to be averaged
over the initial state as well as the operator of the number of
photons (see the example in Appendix \ref{app:colrev}). The
equations Eqs.~(\ref{photon_0order}) and (\ref{photon2_0order}) are
derived in the assumption that the initial state of atoms have zero
dipole momentum. But the inversion in the atomic system is allowed
to have nonzero values. Our initial state satisfies the condition
and the comparison of second order and zero order calculations is
provided in Fig.~\ref{fig:spec5} for $Q(t)$. At short time scales
the agreement is very good.

\begin{figure}
\includegraphics[width=80mm]{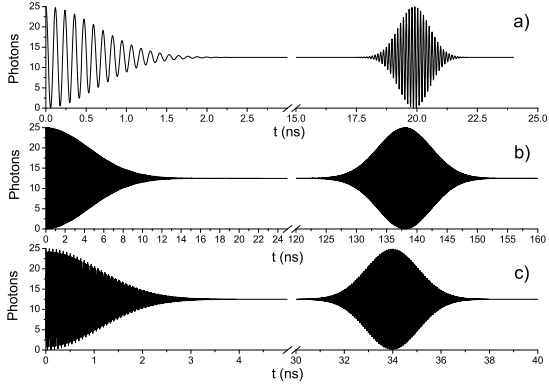}
\caption{\label{fig:spec6} The average number of photons for three
different combinations of the coupling constants $k_1, k_2$. All
parameters are the same as in Fig.~\ref{fig:spec4}. The coupling
constants and the characteristic time scales are given in
Table~\ref{tab:table1}.}
\end{figure}

In a similar way as it was done by Eberly and co-workers
\cite{Eberly:1980}, we calculate the characteristic times of
collapses and revivals in the system by the saddle-point approach.
The details are provided in Appendix \ref{app:colrev}. The revival
time $\tau_{rev}$ is determined by the difference between two
adjacent Rabi frequencies $\Omega_R$ with $M=n_0$ and $M=n_0+1$.
The collapse time $\tau_{col}$ has two different asymptotics. In
the case of very small $k_2$
\begin{eqnarray}
\label{t_col1} \tau_{col}^{(1)}\approx\sqrt{\frac{4(3r-m_0+1)k_1^2
+\Delta^2 }{k_1^4 \sqrt{n_0}}}
\end{eqnarray}
and for relatively large nonlinearity but small $1/r$
\begin{eqnarray}
\label{t_col2} \tau_{col}^{(2)}=\frac{1}{k_2 \sqrt{\sqrt{n_0} r}}.
\end{eqnarray}
In Fig.~\ref{fig:spec6} we demonstrate a good agreement between
these zero order estimates for different ratios $k_2/k_1$, and the
numerical results. It is worth to notice that according to
Eq.~(\ref{t_col2}) the collapse time decreases as $k_2^{-1}$
allowing to achieve significant squeezing before the relaxation
processes take place. As we showed the nonlinear term provides
additional control on the time-scale of quantum effects observed
in the system.

\begin{table}
\caption{\label{tab:table1}Characteristic time scales for three
different combinations of the coupling constants in sodium BEC.}
\begin{ruledtabular}
\begin{tabular}{ccccccc}
   & $k_1/\omega$   & $k_2/\omega$   & $T_R$(ns) & $\tau_{col}^{(1)}$ (ns)
   & $\tau_{col}^{(2)}$ (ns) & $\tau_{rev}$ (ns)\\
 a)& $3\mathrm{E\!-\!7}$ & $-3\mathrm{E\!-\!9}$ & 0.12 & 41.1. & 2.1 & 19.9 \\
 b)& $3\mathrm{E\!-\!7}$ & $-3\mathrm{E\!-\!10}$& 0.10 & 41.1 & 20.8 & 138 \\
 c)& $3\mathrm{E\!-\!6}$ & $-3\mathrm{E\!-\!10}$& 0.01 & 4.1 & 20.8 & 34 \\
\end{tabular}
\end{ruledtabular}
\end{table}

\section{Conclusion}
\label{sec:conclusion}

In the paper, we studied different mechanisms of the nonlinear
interaction between multi-photon optical pulses and many-particles
BEC. The dynamics corresponds to a regime of giant delays in the
probe pulse propagation with almost negligible absorption of the
pulse in the medium with electro-magnetically induced transparency.
Taking the advantage of using tree-level $\Lambda$-scheme within
polarization approach we find an explicit form of the linear and
nonlinear susceptibilities of the atomic condensate characterized
with significant transparency induced by the classical field. We
studied properties of the condensate refraction index and absorption
coefficient related to the probe pulse at frequencies close to the
frequency of corresponding atomic transition. It was demonstrated
that the nonlinear refraction index can reach extremely large
values. The regions of its negative values and nonlinear dependence
of the absorption on the pulse intensity provide a possibility to
achieve almost complete compensation of the losses and the
dispersion in the BEC.

Applying the formalism of polynomially deformed su(2) algebras we
analyze complex quantum dynamics of the probe pulse in a coherent
ensemble of sodium atoms. It was shown that the nonlinear effects
provide a necessary window in the relaxation mechanisms where the
probe pulse in the regime of "slow" light exhibits nonclassical
properties in statistics of photons. We demonstrated that the
poissonian statistics of photons in the coherent state of the probe
pulse can be significantly squeezed within short period of time to
highly subpoissonian values due to collective interaction between
coherent atoms and the field. These results open new perspectives in
generation of nonclassical atomic or field states in Bose-Einstein
condensate controlled by external classical light.

\begin{acknowledgments}
I.V. acknowledges the support of the Engineering and Physical
Sciences Research Council. A.P. acknowledges the financial support
from the Ministry of Education of Russian Federation through Grant
of President of Russia and RFBR (Russian Found of Basic Research)
through Grant No. 01-02-17478. A.P. thank G.Leuchs and N.Korolkova
for many stimulating discussions.
\end{acknowledgments}

\appendix

\section{\label{app:diag} Unitary transformations}
After the transformation $U_0=e^{i\psi_0 \tilde S_y}$ the
Hamiltonian Eq.~(\ref{Hnew}) reads
\begin{widetext}
\begin{eqnarray}
\label{H0_app} &U_0 H^{(2)}U_0^{-1}=C_0+\Omega_R \tilde S_3
-k\frac{\beta_1}4\left[{\cos(2\psi_0)\left({ \tilde S_3 \tilde S_x
+ \tilde S_x  \tilde S_3} \right)+\sin(2\psi_0)\left({ \tilde
S_3^2- \tilde S_x^2} \right)}\right] \nonumber\\
&-k\frac{\beta_2}2\left[{\begin{array}{c}
  (\cos(3\psi_0)+\cos(\psi_0)\sin(\psi_0)^2) \tilde S_3\tilde S_x \tilde S_3
-(\sin(3\psi_0)-\sin(\psi_0)\cos(\psi_0)^2)
\tilde S_x \tilde S_3 \tilde S_x \\
  + \sin(\psi_0)\tilde S_3 \left({ \cos(\psi_0)^2 (\tilde S_3^2-1)
  +\frac14} \right)+ \cos(\psi_0)\tilde S_x \left({ \sin(\psi_0)^2 (\tilde
  S_x^2-1) +\frac14} \right) \\
\end{array} }\right].
\end{eqnarray}

To diagonalize the operator Eq.~(\ref{H0_app}) in first order of
$\beta_1\sim\alpha$ we apply second unitary transformation
\begin{eqnarray}
\label{U1_app} &U_1=\exp{\left[{-i\frac{\beta_1}4\sin(\psi_0)
\left({ (\tilde S_3 \tilde S_y+\tilde S_y \tilde S_3)\cos(2\psi_0)
  -\frac
14 (\tilde S_x \tilde S_y+\tilde S_y \tilde S_x)\sin(2\psi_0)
}\right) }\right]}.
\end{eqnarray}
It is obvious that the transformation $U_1$ applied to the zero
order term in Eq.~(\ref{H0_app}) gives the first order term in the
original Hamiltonian but with an opposite sign providing that it is
cancelled. Thus, the Hamiltonian $U_1 U_0 H^{(2)}U_0^{-1} U_1^{-1}$
is diagonal up to second order of $\alpha$. It has the form
\begin{eqnarray}
\label{H1_app} &U_1 U_0 H^{(2)}U_0^{-1} U_1^{-1}=C_0+\Omega_R
\tilde S_3 -k\frac{\beta_1}4\sin(\psi_0)\cos(\psi_0)\left({3
\tilde
S_3^2 -\tilde r(\tilde r+1)} \right) \nonumber\\
&+k\left({\frac{\beta_1}4}\right)^2\sin(\psi_0)\left[{\begin{array}{c}
  \cos(2\psi_0)^2\tilde S_3(4\tilde S_3^2 -2\tilde r(\tilde
r+1)+\frac12) +\frac{\sin(2\psi_0)^2}4\tilde S_3(5\tilde S_3^2
-5\tilde r(\tilde r+1)+\frac{11}2)\\
+ 3\sin(2\psi_0)^2\tilde S_x \tilde S_3 \tilde S_x
-\frac{\sin(4\psi_0)}2\left({ 9 \tilde S_3 \tilde S_x \tilde
S_3+\tilde S_x^3-
\frac{6\tilde r (\tilde r+1)-13}4 \tilde S_x}\right) \\
\end{array}
}\right]\nonumber\\
&-k\frac{\beta_2}2\left[{\begin{array}{c}
  (\cos(3\psi_0)+\cos(\psi_0)\sin(\psi_0)^2) \tilde S_3\tilde S_x \tilde S_3
-(\sin(3\psi_0)-\sin(\psi_0)\cos(\psi_0)^2)
\tilde S_x \tilde S_3 \tilde S_x \\
  + \sin(\psi_0)\tilde S_3 \left({ \cos(\psi_0)^2 (\tilde S_3^2-1)
  +\frac14} \right)+ \cos(\psi_0)\tilde S_x \left({ \sin(\psi_0)^2 (\tilde
  S_x^2-1) +\frac14} \right) \\
\end{array} }\right].
\end{eqnarray}
Notice that we only keep terms of order $\alpha$ or $\alpha^2$.
Therefore, the last term in Eq.~(\ref{H0_app}) stays intact after
the transformation. One can see that the first order correction
vanishes in the case of exact resonance $\Delta=0$. So, it is
necessary to consider second order terms to take into account
significant effects connected with the nonlinear dependence of the
energy splitting on the value of collective atom-field quasi-spin
vector $\tilde S_3$.

To diagonalize the operator Eq.~(\ref{H1_app}) up to the order
$\alpha^2$ we apply the third unitary transformation
\begin{eqnarray}
\label{U2_app}
&U_2=\exp{\left[{\begin{array}{c}i\left({\frac{\beta_1}4\sin(\psi_0)}\right)^2
\left\{{
  \frac34\sin(\psi_0)^2(\tilde S_x \tilde S_3 \tilde S_y
+\tilde S_y \tilde S_3 \tilde S_x)
  +\frac{\sin(4\psi_0)}2\left({\frac13 \tilde S_y^3-8\tilde S_3 \tilde
  S_y\tilde S_3+\frac{\tilde r(\tilde r+1)}2\tilde
  S_y-\frac{31}{12}\tilde S_y}\right)}\right\} \\
  -i\frac{\beta_2}2\sin(\psi_0) \left\{{
  \begin{array}{c}
  2\cos(3\psi_0)\tilde S_3\tilde S_y \tilde S_3+\frac12\cos(\psi_0)\tilde S_y
  + \frac12\sin(\psi_0)\cos(\psi_0)^2(\tilde S_y \tilde S_3 \tilde S_x+
  \tilde S_x \tilde S_3 \tilde S_y)\\
 -\frac{\sin(3\psi_0)}2
  (\tilde S_y \tilde S_3\tilde S_x+\tilde S_x\tilde S_3\tilde S_y
  ) - \cos(\psi_0)\sin(\psi_0)^2\tilde S_y(\frac23\tilde S_y^2-2\tilde
  r(\tilde r+1)+\frac{10}3) \\
\end{array}
}\right\}
\\
\end{array}}\right]}.\quad
\end{eqnarray}
\end{widetext}
Consecutively applying these three transformations $U_{0,1,2}$ to
the Hamiltonian Eq.~(\ref{Hnew}) we obtain the diagonal operator
Eq.~(\ref{eq36}).

\section{\label{app:colrev} The collapse and revival characteristic times }
According to the zero-order solution Eq.(\ref{photon_0order}) a
dynamics of the average number of photons for the initial state
Eq.(\ref{init_st}) has the form
\begin{eqnarray}
\label{nav_app} \bar n(t)=\sum\limits_{n=0}^\infty\frac{n n_0^n
e^{-n_0}}{n!}
\frac{k_n^2+2\Delta^2+k_n^2\cos(\sqrt{k_n^2+\Delta^2}t)}
{2(k_n^2+\Delta^2)} ,\quad
\end{eqnarray}
where based on Eqs.~(\ref{params1}) and (\ref{params2}) we defined
\begin{eqnarray}
\label{kn_app} k_n=\left({ k_1+k_2\frac{n+1}2}\right)
\sqrt{2(4r-n+1)}.
\end{eqnarray}
Deriving the expressions we used the fact that our initial state
belongs to the nearby zones ($\tilde r=M/2$) and the atoms are
completely unexcited ($M=n, m_0=-r$). For simplicity of the
analytical expressions below, we represent the results taking
$m_0=-r$. But they can easily be generalized for arbitrary initial
number of excitations in atomic subsystem. Since we are interested
in the time dependence of $\bar n(t)$, the constant terms in
Eq.~(\ref{nav_app}) can be discarded. We estimate the sum of the
time depending term using the saddle-point method, which has been
applied to calculate the one-atom model \cite{Eberly:1980}. Denoting
the time dependent term in Eq.~(\ref{nav_app}) by $w(t)$ we can
write
\begin{eqnarray}
\label{nav1_app} w(t)\approx\int\limits_{0}^\infty\sqrt{\frac
n{2\pi}} \frac{k_n^2} {2(k_n^2+\Delta^2)}Re\left({e^{-n_0+n_0
f(n/n_0)}
 }\right)\,dn,\quad
\end{eqnarray}
where
\begin{eqnarray}
\label{f_app} &f(x)=x(1-\ln x)\nonumber \\
&+it \sqrt{ \frac{\Delta^2} {n_0^2} + \left({2r+\frac12-\frac{n_0
x}2}\right) \left({\frac{2k_1+k_2}{n_0}+k_2 x} \right)^2}.
\end{eqnarray}
Using the approach we find
\begin{eqnarray}
\label{nav2_app} w(t)\approx  \frac{n_0 k_{n_0}^2}
{2(k_{n_0}^2+\Delta^2)} Re \left({\frac{e^{-n_0+n_0 f(x_0)}
}{\sqrt{f''(x_0)}} }\right),
\end{eqnarray}
where the point $x_0$ is the saddle-point of the analytical function
$f(x)$, i.e. $f'(x_0)=0$. It follows from the definition that the
saddle-point $x_0$ depends on the time. As it was explained in
detail in the paper \cite{Eberly:1980}, the collapse time
$\tau_{col}$ is roughly estimated by the condition
\begin{eqnarray}
\label{tcol_con} \left|{1-Re f(x_0)}\right|=\sqrt{n_0}.
\end{eqnarray}
For the moment $\tau_{col}$ when the condition
Eq.~(\ref{tcol_con}) is fulfilled, the exponent in
Eq.~(\ref{nav2_app}) becomes very small and the envelope of the
sinusoidal oscillations "collapses". For $t=0$ it is plain to see
that $x_0=1$.

Expanding the solution for $x_0$ in the vicinity of unity and
substituting it into Eq.~(\ref{tcol_con}) we obtain relatively
lengthy expression. If we assume that the nonlinear susceptibility
$k_2$ is the dominating smallness parameter, and expand the result
with respect to $k_2\rightarrow0$, we find
\begin{widetext}
\begin{eqnarray}
\label{tcol1_app}
\tau_{col}^{(1)}\approx\sqrt{\frac{4(3r-m_0+1)k_1^2 +\Delta^2
}{k_1^4
\sqrt{n_0}}+4k_2\frac{k_1^2(8r-n_0+1)(4r+1)-(2r-n_0)\Delta^2}{k_1^5\sqrt{n_0}}}.
\end{eqnarray}
Assuming that the leading smallness parameter is $1/r$ we obtain an
expression for relatively large nonlinearity $k_2$
\begin{eqnarray}
\label{tcol2_app} \tau_{col}^{(2)}=\frac{1}{k_2 \sqrt{\sqrt{n_0}
r}}\sqrt{1+\frac{8k_1^4+6k_1^2k_2^2(2n_0+1)+4k_1^3k_2(2n_0+3)
+k_1k_2^3(\frac{\Delta^2}{k_2^2}+6n_0+1)+k_2^4(n_0-\frac{\Delta^2}
{k_2^2}(n_0-\frac12))}{2rk_2(2k_1+k_2)^3}}.\quad
\end{eqnarray}
\end{widetext}
Even for $k_2=0$ in the Eq.~(\ref{tcol1_app}) our results can't be
straightforward compared with Eberly's paper because we consider the
dynamics in the nearby zones, which do not exist in the simple
one-atom model. But, it is not a problem to obtain corresponding
results for the remote zones, which are more relevant to their work.
We also want to notice that the solutions Eq.~(\ref{tcol1_app}) and
(\ref{tcol2_app}) provide additional information about the
dependence of collapse times on the number of particles in BEC,
which is completely opposite in these two asymptotics.

The revival time is estimated by the period when the phases of
oscillations of neighboring terms in Eq.~(\ref{nav_app}) with
frequencies $k_n$ differ by $2\pi$. The difference is estimated for
the dominant sinusoids with $n=n_0$ and $n=n_0+1$. The revival time
reads
\begin{eqnarray}
\label{trev_app} \tau_{rev}\approx\left|{
\frac{2\pi}{\sqrt{k_{n_0+1}^2+\Delta^2}-\sqrt{k_{n_0}^2+\Delta^2}}}\right|.
\end{eqnarray}
The Rabi oscillation period $T_R$ in Table~\ref{tab:table1} is
estimated as
\begin{eqnarray}
\label{T_R_app} T_R\approx \frac{2\pi}{\sqrt{k_{n_0}^2+\Delta^2}}.
\end{eqnarray}
\newpage

\bibliography{paper}

\end{document}